\documentclass[aps,preprint,superscriptaddress,showpacs,showkeys]{revtex4}
\usepackage{graphics,graphicx,dcolumn,bm,fleqn,epic,eepic,float}
\usepackage{amssymb,amsmath,multirow,rotate,color,float}
\usepackage{setspace}
\begin{document}
\title{Community structure and ethnic preferences in school friendship networks}
\author{M. C. Gonz\'alez}
\affiliation{Institute for Computational Physics, 
             Universit\"at Stuttgart, Pfaffenwaldring 27, 
             D-70569 Stuttgart, Germany}
\affiliation{Departamento de F\'{\i}sica, Universidade Federal do
             Cear\'a, 60451-970 Fortaleza, Brazil}
\author{H.J. Herrmann} 
\affiliation{IfB, HIF E12, ETH H\"{o}nggerberg, CH-8093 Z\"{u}rich, Switzerland}
\affiliation{Departamento de F\'{\i}sica, Universidade Federal do
             Cear\'a, 60451-970 Fortaleza, Brazil}
\author{J. Kert\'esz}
\affiliation{Instiute of Physics, 
             Budapest University of Technology and Economics, 
             H-1111 Budafoki \'{u}t. 8., Budapest, Hungary} 
\author{T. Vicsek}
\affiliation{Biological Physics Ressearch Group of HAS,
             E\"{o}tv\"{o}s Lor\'{a}nd University,
             H-1117, Pazmany P. S\'et\'any 1A, Budapest, Hungary} 
\date{\today}

\begin{abstract}
  
  Recently developed concepts and techniques of analyzing complex
  systems provide new insight into the structure of social networks.
  Uncovering recurrent preferences and organizational principles in
  such networks is a key issue to characterize them. We investigate
  school friendship networks from the Add Health database. Applying
  threshold analysis, we find that the friendship networks do not form
  a single connected component through mutual strong nominations
  within a school, while under weaker conditions such
  interconnectedness is present. We extract the networks of
  overlapping communities at the schools (c-networks) and find that
  they are scale free and disassortative in contrast to the direct
  friendship networks, which have an exponential degree distribution
  and are assortative. Based on the network analysis we study the
  ethnic preferences in friendship selection. The clique percolation
  method we use reveals that when in minority, the students tend to
  build more densely interconnected groups of friends. We also find an
  asymmetry in the behavior of black minorities in a white majority as
  compared to that of white minorities in a black majority.
\end{abstract}
\maketitle
Social structures in schools are subject to intense investigations for many obvious reasons. Schools visited by major part of the population form social systems, which are well defined units enabling to study relationships, networking and processes in a condensed way.  The relationships of adolescents show remarkable peculiarities, they are influenced by family backgrounds and, at the same time, they are precursors of the future society. The problems of spreading sexually transmitted diseases, of drug abuse or of delinquency among adolescents and young adults are closely related to their social embedding in the schools and so are their racial/ethnic preferences.

The investigation of patterns of friend selection is a major source of our knowledge on social structures in schools \cite{Epstein}. Mapping out the friendship networks based on questionnaires have been a successfull approach in this respect, where the existence and intensity of dyadic connections are identified using nominations of the students~\cite{Hansell, Brewer, Baerveldt1, Moody}. It is known that sex and race/ethnicity are two primary characteristics on which students base their selections of friends \cite {Epstein} and here we would like to focus on the latter. 

Desegregation of schools as a function of the racial diversity has been a topic of analysis in multi-ethnic countries in  Western Europe~\cite{Baerveldt1, Baerveldt2} and the USA~\cite{Joyner, Moody, Kao}. These studies suggested that the way schools are organized could affect the level of racial friendship segregation. In recent studies of friendship networks $p*$ and related models \cite {Wassermann, Wassermann2} were successfully used to identify how some of the attributes of the network members are correlated with their inclinations in choosing group relationships. However, as the measures of segregation are still under discussion \cite {Gorard},and even racial classification schemes seem problematic \cite{Harris}, we think it useful to approach this problem from a different angle, namely to apply concepts and results from the science of complex networks \cite{Albert, Newman_siam, Dorogovtsev}.

These results include the quantitative characterization of hierarchical ordering \cite{Ravasz_hier, Sneppen_hier}, new, efficient methods of community detection \cite{Newman_comm, Bornholdt, Castellano, Derenyi}, even of overlapping ones \cite{Palla} and pointing out relations between functionality and weights of the links in the network \cite{Barrat, Onnela}. Successful efforts have been made to analyze complex networks, including social ones, within this new framework ~\cite{Colizza,GonzalezPRL,Porter,Alon,Sneppen,Amaral}. 

Our aim here is to present an analysis of friendship networks in schools based on the representative US National Longitudinal Study of Adolescent Health (Add Healt, \cite {AddHealth}). First we carry out a topological study  and apply threshold analysis \cite{Onnela_threshold} in order to identify the network which is most appropriate for our further investigation. In contrast to earlier work, our study focuses on the communities instead of the dyadic links. Interestingly, we uncover that the properties of the direct friendship network are significantly different from the network of the next hierarchical level, namely the network of communities.

\section*{Friendship networks}
\label{first_view}    
The friendship networks presented here are constructed from the in-school questionnaires of the Add-Health~\cite{AddHealth} friendship nomination study from the period 1994-1995, in which 90118 students participated. The analyzed data are limited to students who provided information from schools with response rates of $50 \%$ or higher. Every student was given a paper-and-pencil questionnaire and a copy
of a list with every schoolmate. Weighted dyadic links were generated based on the number of sheared activities. Weights were in the range from $1$, meaning the student nominated the friend without reporting
any activity, to $6$ meaning that the student nominated the friend and reported participating in all five activities with (him/her). 

The structure files contain information on $75871$ nodes divided in $84$ networks (schools). 
In most of the analyzed samples of schools the majority of the population is white, however, there are significant fluctuations. In particular, the ratios of the races in the total population is the foolowing: White:0.59, Black:0.14, Hispanic:0.13, Asian:0.04 and Other:0.1.
\begin{figure*}[htb]
\begin{center}
\includegraphics[width=16.0cm]{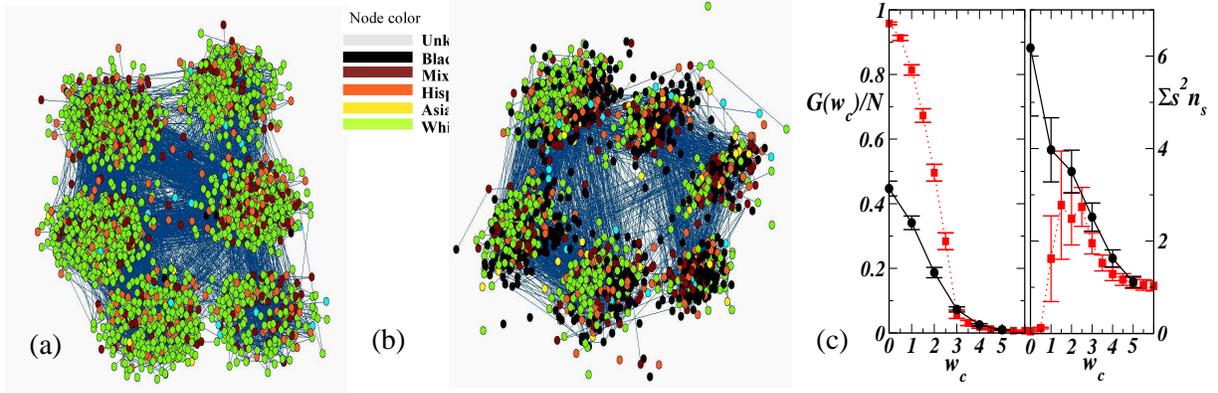}
\end{center}
\caption{\protect {\bf(a)-(b)} Networks of friendships from Schools $1$ and $2$ (respectively). Nodes represent students, with colors indicating their race. Spatial distribution of nodes corresponds to the different grades,
placed counter clockwise, from 7th to 12th grade.{\bf(c) Left:} $G/N$ fraction of sites in the largest connected component $G$ for the networks with mutual links only (circles) and networks with mutual and not mutual links (squares) versus threshold weight $w_{c}$. Only links with average weight in both directions $w \ge w_{c}$ are kept. {\bf Right:} Second moment of the normalized number of clusters excluding the largest component for the same analysis as in the left part.}
\label{fig1}
\end{figure*}
In Figs.~\ref{fig1}a-b, we visualize the friendship networks for two schools with pajek\cite{pajek}. 
Fig.~\ref{fig1}a is a characteristic sample of the $84$ schools, we call it here School $1$. In this school the great majority of the population is white ($70\%$), which contrasts to a non-characteristic  sample, School $2$, visualized in Fig.~\ref{fig1}(b), where blacks ($40\%$) are overrepresented with respect to the average.
Nodes represent students, with colors indicating their race. A link is drawn between nodes if at least one of the student nominates the other like a friend. The spatial distribution of the nodes corresponding to the different grades, placed counter clockwise, starting with the 7th grade at lower right corner and ending with the 12th grade. Visual inspection of the intergrade links already tells that there is a separation between the upper grades (high school) and the lower grades (middle school). While the partition according to the grades was introduced "by hand" the separation of colors within the $6$ groups is not artificial; the apparent clustering of nodes according to the same color is due to the fact that they are more densely interconnected.
 
\section*{Role of weights and directionality} 

Checking mutuality in a whole-network study \cite{Marsden} gives some insight into the reliability of the answers given to the questionnaires. In an ideal case both participants of a dyadic relationship should name each other with the same weight. We apply threshold analysis to measure the influence of weights and directionality in the links. In order to analise the role of the weights we take an average over all schools.

First, we analyze the network formed only by mutual links, i.e., mutual nominations, which should have the more reliable information about, stronger relations or tight friendships inside the networks. We introduce the mean of the weight in both directions to characterize the weight of each link ($w$). We examine different thresholds of ($w_{c}$) for creating links, i.e., a link is created only if there is a mutual link and $w \ge w_{c}$. The values of the weights go from $1$ to $6$, the weakest possible restriction is $w=1$, which includes any mutual link present in the network. In the left part of figure ~\ref{fig1}c (circles) we present  the  calculations of $G/N$, the fraction of nodes that belong to the largest cluster vs. $w_{c}$. On the right side, $\sum_{s} s^{2}n_{s}$, the second moment of the normalized number of clusters $n_s$ of size $s$ (excluding the largest cluster) is presented. Interestingly, when considering only mutual connections $G$ is roughly half of the population, and the network is split in various components.

Next we make the threshold analysis by considering the network as follows: A link is formed if at least one nomination exists, and ($w > w_{c}$); the weight $w$ of a link is taken again as the mean of the weights in both directions with the extension that for the direction into which the nomination does not exist, zero is taken. For this case, we find a transition as a function of $w_c$: The population is disconnected into many clusters for $w_c > 2$ while a giant component occurs for $w_c \le 2$. This effect is shown on both sides of Figs.\ref{fig1}c (red squares). We have found that only the weakest threshold criterium and dropping the requirement of mutuality leads to a spanning giant component. This finding harmonizes with the
finding~\cite{Cairns} that applying strong criteria for constructing friendship networks leads to a network instability while with weak criteria the network turns out to be stable.

In our further analysis of community detection we assume that a dyadic link exists if any of the corresponding students nominates the other, and we do not consider any threshold for the weight. Imposing the minimum restriction possible for the creation of a link allows us to search for communities in the interconnected giant component and to uncover preferences in the social relationships between the students.

\section*{Networks of communities (c-networks)}
\begin{figure*}[tb]
\begin{center}
\includegraphics[width=16.0cm]{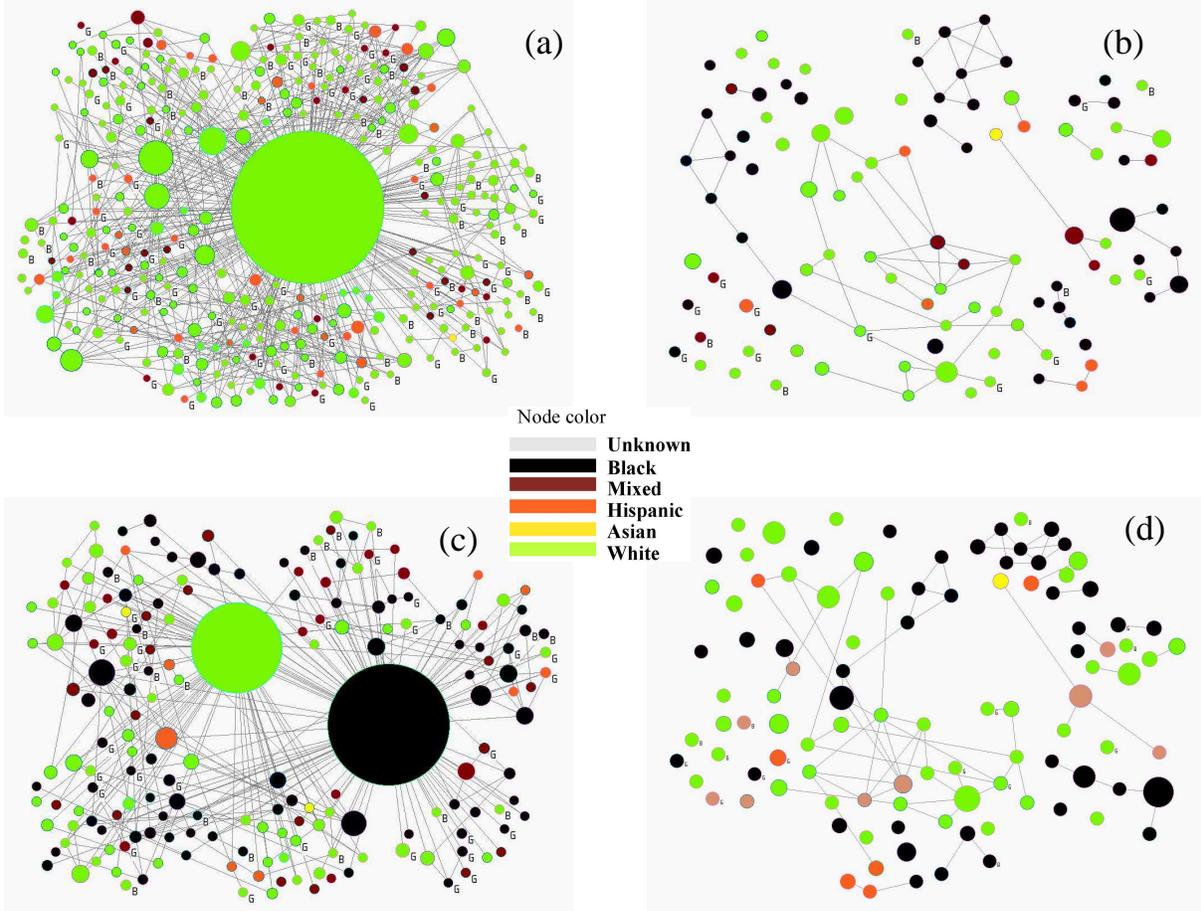}
\end{center}
\caption{\protect C-networks of 3-clique communities 
at School 1 ({\bf(a)}) and School 2 ({\bf(c)}).
Compared to the corresponding c-networks
of 4-clique communities ({\bf (b)} and {\bf (d)} respectively). 
The color is assigned according to the race of the majority of 
nodes in the community. The node size is proportional to the square 
root of the number of nodes in the community. Although, each community can have students from different races, we assign to it the color of the majority of the members of the community.} 
\label{fig2}
\end{figure*}

The social network reflects the structure of the society. Therefore it carries information about the building bricks, the communities. However, it is a highly non-trivial task to extract this information from the network itself. Communities are vaguely defined as groups of vertices that have a high density of edges within them, with a lower density of edges between the groups~\cite{Scott,Everitt}. The recently introduced method of community detection, the ``clique percolation method''~\cite{Palla, Derenyi} seems particularly appropriate to handle this problem because it enables overlapping communities, which are typical for the social networks. Two communities overlap, if they share at least one member. In most of the friendship groups there are members, who simultaneously belong to more than one such group. This feature is known as "affiliation" (see, e.g., \cite{Wassermann} in the social networks literature and is an aspect of large networks which is on one hand very important, while it has not been satisfactorily addressed by the recently developed (prior to the k-clique percolation approach) network clustering methods.

A {\it $k$-clique} is a fully connected subgraph containing $k$ nodes. A {\it $k$-clique community} is defined as a group of $k$-cliques that can be reached from each other through a series of adjacent $k$-cliques sharing $k-1$ nodes. After determining the $k$-clique communities, it turns out that there are nodes which belong to more than one community. Using these shared nodes one can construct the c-network of communities, where the communities themselves constitute the c-nodes and the shared nodes of the original network form the c-links between them. In the following we analyze the c-network of communities based on the friendship networks of the schools.

Fig.~\ref{fig2} shows the c-network of $k$-clique communities extracted from the friendship networks of school $1$ and school $2$. Figs.~\ref{fig2}a and c is based on $3$-clique communities of friendship networks of School $1$ (Fig.~\ref{fig1}a) and school $2$ (Fig.~\ref{fig1}c), respectively. In turn, Figs.~\ref{fig2}b and d are based on $4$-clique communities extracted from the same schools. The area of the circles represents the number of nodes within the community and each node color is related to a race.

A comparison of Figs 2a-c with Figs 2b-d shows that there is a dramatic difference between the c-networks based on $3$-clique or $4$-clique communities. For the $3$-clique communities we see in both schools complex c-networks with rich, interconnected structures, which include the great majority of the students, while the c-networks of $4$-clique communities are rather sparse (less than $20\%$ of the students belong to them) and the structures are fragmented. 

It has been suggested~\cite{Palla} that the optimal value of $k$ for uncovering the community structure in a network is the largest one which still assures percolation, i.e., interconnectedness.
In contrast to other studied networks~\cite{Palla}, like protein networks or collaboration networks, where the optimal value for detecting communities was $k=4$ or $5$, we have found that triads are the optimal elementary cliques for the high school friendship networks. Although it is shown here only for schools 1 and 2, our finding is generally valid for the whole data set. 
This is a new manifestation of the well known fact that triads play an eminent role in interpersonal relations \cite{Heider, Szvetelszky}, which is also reflected in the high value of the average clustering coefficient \footnote{The clustering coefficient of a node is the ratio of the linked neighboring nodes to the total number of neighbor pairs. This quantity will be discussed later in the paper.} of social networks\cite{Watts}. 

Although we obtain the richest community structure for $k=3$, it is worth having a look at the c-networks based on the more cohesive $4$-cliques. 
For $k=3$ already the relatively less densely connected friendship circles show up in the analysis, while for $k=4$ only the more strongly interconnected groups (in which each member is part of at least one 4-clique) are found by the method. 
One of the interesting aspects of such a study is that on the level of 
more cohesive groups ($k=4$) the number of communities becomes balanced even for cases when the ratio of the sizes of the ethnic groups is far from unity (and, correspondingly, on the level of less cohesive groups, e.g., for $k=3$, the students who are in majority, have much larger friendship circles). From here (see Figs.~\ref{fig2}b and d) we conclude that when in minority, the students tend to form stronger ties, thus, the number of more densely interconnected communities becomes over-represented compared to what happens in the $k=3$ case. 

\begin{figure*}[tb]
\begin{center}
\includegraphics[width=16.0cm]{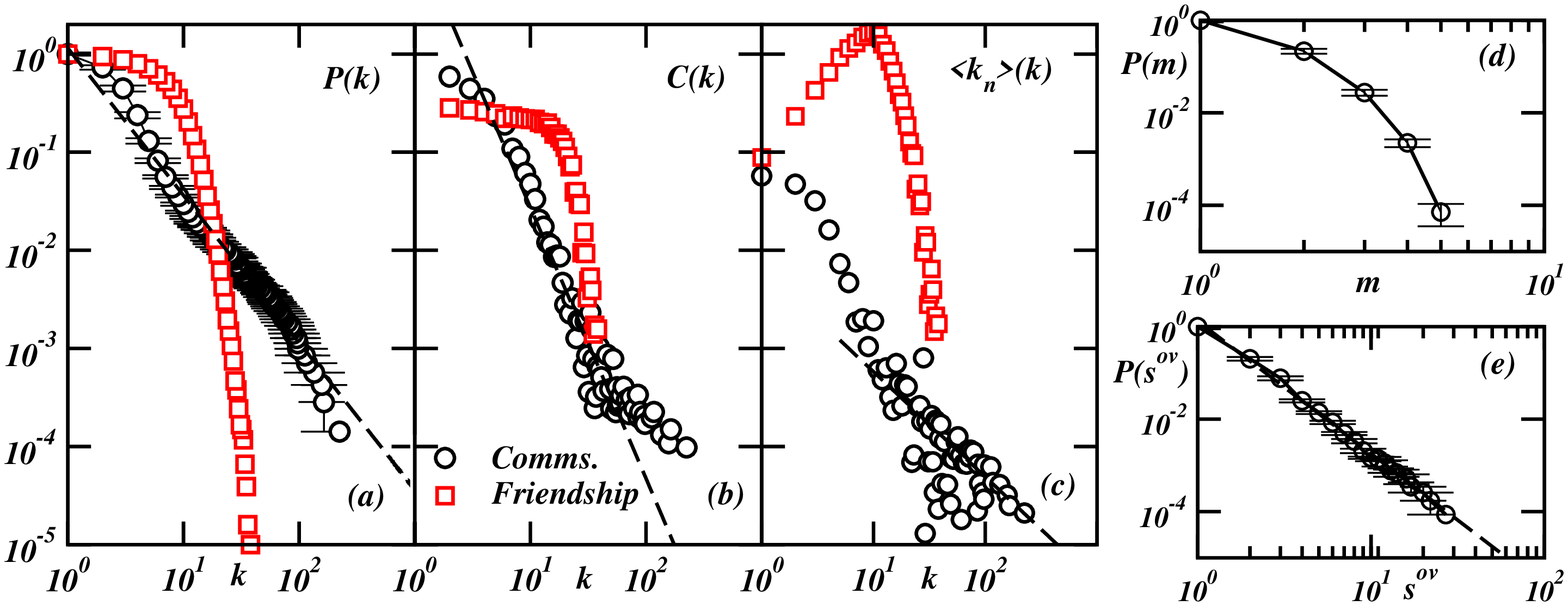}
\end{center}
\caption{\protect Different network properties averaged over the complete
  dataset of schools,
for the community networks (circles) and for friendship networks 
(squares): {\bf (a)} Cumulative degree distribution.
{\bf (b)} Degree-dependent clustering coefficient. {\bf (c)} Average
degree of the nearest neighbor. {\bf(d)} Cumulative distribution
of the membership number ($m$) and of {\bf(d)} the overlap size ($s^{ov}$)
for the community networks.} 
\label{fig3}
\end{figure*}

\section*{Statistical properties of the c-networks}

In the following we statistically characterize the structure of the friendship networks and of the extracted c-networks based on $3$-clique communities, where averages will be taken over all 84 schools in the data set.

The cumulative degree distribution $P(n)$ is defined as the fraction of nodes having degree larger than $n$. In Fig.~\ref{fig3}a we show $P(n)$ for the friendship networks and compare it with the cumulative degree distribution of the c-networks of communities. The distribution for the friendship networks rapidly  decreases, indicating that these networks have a characteristic degree. This corresponds to the natural cutoff in the number of friends, in accordance with the results reported for another friendship network~\cite{Amaral}. Interestingly, the degree distribution of the c-networks is much broader, and can be well fitted by a scale free, power-law function of the form $\sim n^{-\gamma}$ with $\gamma \approx 1.5$. It is known that such {\it scale free networks} emerge from growth processes where an effective preferential attachment, i.e., a "rich get richer" mechanism is at play \cite{Albert}. Scale free c-networks have already been seen before~\cite{VicsekLett}, but the transition from the rapidly decaying degree distribution in friendship network to the scale freeness of the c-networks is a relevant characteristic of social community formation and should be taken into account for the formulation of models of large social networks~\cite{GonzalezPRL}.

The degree distribution provides information about the dyadic relations while the clustering coefficient characterizes the triads. The local clustering coefficient ($C_{i}$) of a vertex $i$ with degree $n_{i}$, is defined as the ratio of the number of triangles connected to it and all the possible number of triangles ($n_{i}(n_{i}-1)/2$). The mean degree-dependent clustering coefficient is the average of the local clustering over all vertices with degree $n$. This quantity is analyzed for the two types of networks and presented in Fig.~\ref{fig3}b. For the friendship networks $C(n)$ varies slightly with $n$ for most of the observed $n$-range; decaying rapidly only for larger degrees. Again, C(n) for the c-network is much broader than for the friendship network and can be reasonably fitted by a power law $C(n) \sim n^{-\alpha}$, with $\alpha \approx 2.8$. This kind of dependence of the clustering coefficient as an inverse power of the node degree, can be signature of a hierarchical structure of the networks~\cite{Ravasz,Kertesz,Havlin}.

Social networks are known to be assortative, i.e., high degree nodes are linked with enhanced probability.
The statistical analysis of this effect relies on the degree $\tilde {n}(n)$ of nearest neighbors averaged over all nodes of degree $n$. For assortative (disassortative) network $\tilde {n}(n)$ is a monotonously increasing (decreasing) function of $n$. As expected, the friendship networks turn out to be assortative (see Fig.~\ref{fig3}c), but in contrast to networks with scale free degree distribution (e.g., collaboration networks), $\tilde {n}(n)$ has also a 
cutoff due to the rapid decay in the degree distribution. On the other hand, the c-networks are
disassortative, i.e., $\tilde {n}(n)$ can be approximated by a power law with a negative exponent, $\tilde {n}(n) \sim n^{-\beta}$, with $\beta \approx 1.1$. 

We also calculate the membership ($m$) of each student, which is the number of communities that the students belongs to.  Fig.~\ref{fig3}d displays the cumulative distribution of the membership number $P(m)$, which shows that on average, each student belongs to a limited number of communities (less than $5$). In turn, any two communities can share $s^{ov}$ nodes, which defines the overlap size between these communities. Fig.~\ref{fig3}e shows the average of the overlap distribution for all the schools, which is well fitted by a power law with the exponent $2.9$. We can conclude that students belong to at most $4$ different clique-communities inside the School, and that there is no characteristic overlap size in the networks (except of that given by their finite size). Absence of characteristic membership number and overlap size have been observed in other social and biological networks but not in their randomized versions~\cite{Palla}. Additionally, the clustering coefficient, $\langle C \rangle$ for friendship networks and for community networks both have a similar average value 
near $0.3$, which is larger than an equivalent random graph with the same number of nodes and links.

\begin{figure*}[tb]
\begin{center}
\includegraphics[width=16.0cm]{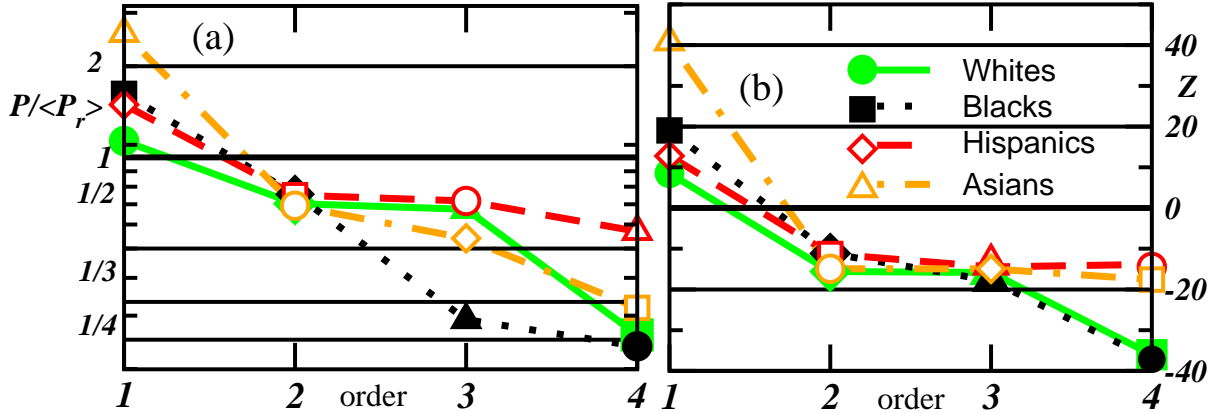}
\end{center}
\caption{\protect
Measuring preferences of inter-racial connections $r-r'$. $P(r,r')$ is the relative frequency of directed links from Whites (full green line), Blacks (dotted black line), Hispanics (dashed red line) and Asians (dashed-dotted yellow line) to each of the races $r'=W$ (circles), $B$ (squares), $H$ (diamonds), and $A$ (triangles). Racial preferences manifest themselves as systematic deviations of the ratio $P(r,r')/P_r(r,r')$
from $1$, $P_r$ is the corresponding relative frequency in the randomized samples. {\bf (a)} $P/P_r$
in decreasing order from $1$ to $4$, for the nominations made from 
$r$ to $r'$. {\bf(b)} The corresponding $Z$-scores. The combination of  {\bf(a)} and {\bf(b)} reveals
relations $r-r'$ that are significantly absent. 
The results are the average over the $84$ School networks.}
\label{fig5}
\end{figure*}

\section*{Ethnic preferences}
Racial/ethnic preferences in friendship selection contain crucial information about the level of segregation, which constitutes one of the major sources of social conflicts. Quantifying such concepts as preferences or segregation and to work out the appropriate measurement protocols are highly non-trivial tasks in a strongly inhomogeneous society (see \cite{Gorard, Harris}).

We use the following quantitative method to measure `{\it preferential}' nominations as a function of the attributes of the students. A nomination can be considered preferential, if pairs of nodes with given attributes 
are significantly more recurrent within the empirical networks than those in their randomized versions. In the studied sample of friendship networks, we find the dominant appearance of quantitatively preferential nominations among students of the same race, as a manifestation of homophily 
present in each grade and common to each racial group 
from all schools. Here we present in detail the measure of preferences 
in the School networks as a function of the race known for the nodes, without separating the information by grade. The same method can, of course, be used to measure preferences in any attributes.

In each directed network we identify  the frequency of the $25$ possible race dyads, formed from the $5$ races attributed to the nodes. To focus on those dyads that are significantly recurrent, we compare the real network to suitably randomized networks.

The randomized networks have the same single node characteristics as those the real networks: Each node in the randomized network keeps its race and the same number of incoming and outgoing edges as the corresponding node has in the real network. For randomizing the networks we employ a Markov-chain algorithm, based on starting with the real network and repeatedly swapping randomly chosen pairs of connections ($A \rightarrow B$, $C \rightarrow D$ is replaced by $A \rightarrow D$, $C \rightarrow B$) until the network is randomized ~\cite{Sneppen,Alon}. Switching is prohibited if either of the connections $A$ and $D$ or $C$ and $B$ already exist. Thus the degree of each node is preserved.

In Fig.~\ref{fig5} we present results for the main $4$ races identified at the schools: white, black, hispanic and asians. For each race $r$, we calculate the relative frequency $P(r,r')$ directed links $r \rightarrow r'$, to a node with race $r'$. The presented results are the average over the $84$ schools. The comparison to randomized networks compensates for the effects of differences in the amount of each race population. 
Racial preferences manifest themselves as systematic deviations of the ratio $P(r,r')/<P_r(r,r')>$ from $1$.
The common behavior for each racial group is to nominate friends of the same race (intra-ethnic nominations) more likely than students from any of the other race (inter-ethnic nominations).  In Fig.~\ref{fig5}a, we present $P/<P_r>$ in decreasing order from $1$ to $4$, for the nominations made for each race $r$ (denoted by different line styles and colors) the race of the nominated nodes $r'$ (indicated by different symbols). Not only the preference for intra-ethnic nominations becomes clear from this plot, but also that symmetrically
some inter-ethnic nominations are found $4$ times less often than in the randomized versions, e.g., those from asians $ \leftrightarrow $ blacks and blacks $\leftrightarrow$ whites. In Fig.~\ref{fig5}b, we characterize the significance of the deviations by the $Z$-scores, defined as: 
\begin{equation}
Z(r,r') \equiv \frac{P(r,r')-<P_{r}(r,r')>}{\sigma_{r}(a,a')}, 
\end{equation}
where $\sigma_{r}(r,r')$ is the standard deviation of $<P_{r}(r,r')>$ calculated from $100$ realizations of randomized networks. The combination of these two plots reveals relations $r \leftrightarrow r'$ that are significantly absent. 

\begin{figure*}[tb]
\begin{center}
\includegraphics[width=16.0cm]{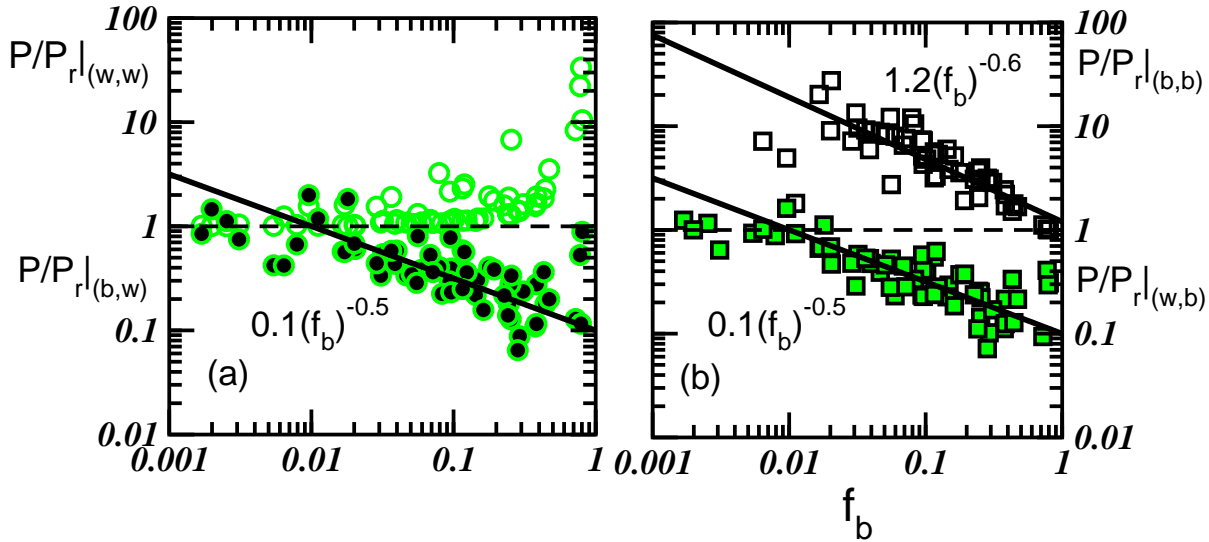}
\end{center}
\caption{\protect
The ratio of the relative frequencies ($P/P_{r}$) vs. fraction of the minority,
i.e. black population ($f_{b}$). {(\bf a)}For $white \rightarrow white$
and $white \rightarrow  black$ nominations. {(\bf b)} For $black \rightarrow  black$ 
and $black \rightarrow white$ nominations.  $P/<P_{r}>$ can be fitted by a negative power law
of the form $f_{b}^{-\alpha}$, with $\alpha = 0.6$. For $black \leftrightarrow black$ nominations 
and $\alpha = 0.5$ for $black \leftrightarrow white$ nominations. This shows that although heterogeneity 
decreases the relative frequency of  $b \leftrightarrow b$ , it does not
favor inter-ethnic relations $b \leftrightarrow w$.}
\label{fig6}
\end{figure*}

Next, we illustrate how the measured quantity $P(r,r')/<P_r(r,r')>$, can be used to obtain certain characteristics
of the friendship selection preference as a function of the racial composition of the schools. In the following, we focus on the relations of two ethnic groups: blacks ($b$) and whites ($w$). In Fig.~\ref{fig6} we represent the obtained value of $P/<P_{r}>$ vs. the fraction of the minority ($f_{b}$), i.e. students of the black population in each school network. Figure ~\ref{fig6}a shows the values for the nominations from whites, intra-ethnic $w \rightarrow w$ and inter-ethnic $w \rightarrow b$. Equivalently, Fig.~\ref{fig6}b shows the corresponding nominations from blacks $b \rightarrow b$ and $b \rightarrow w$. These figures show a sample of $64$ schools which have at least $0.2\%$ of any of both races (white and black).

We have observed that intra-ethnic nominations occur equally or more frequently than in the randomized networks
($P/<P_{r}>\ge 1$), while inter-ethnic nominations are less likely to occur ($P/<P_{r}> < 1$), and these results do not depend on the total size of the population, $N$ (not shown). When we plot the same quantities as a function of the fraction of the minority it is possible to extract some relevant tendencies from the entire sample. Note that $P/<P_{r}>$  vs. $f_{b}$, for $b \rightarrow b$ is greater than $1$ and tends to $1$ only when $f_{b} \sim 1$ (top of Fig.~\ref{fig6}b), just for such values $P/<P_{r}>$ of $w \rightarrow w$ is then considerably greater than $1$ (top of Fig.~\ref{fig6}a). These figures show that both races present the following behavior: When the population of a given race, is majority (fraction $f \sim 1$), then their intra-ethnic nominations resemble those of the randomized networks $P/<P_{r}> \sim 1$, but when they represent a minority ($f \ll 1$) such populations tend to make intra-ethnic nominations of friends ($P/<P_{r}> \gg 1$). 

In contrast to the intra-ethnic relations, the inter-ethnic nominations are non-symmetric with respect to the composition. This is clearly shown in~\ref{fig6}a) and b). It is natural that the $b \rightarrow w$ and $w \rightarrow b$ follow the same pattern. However, in case of symmetric behavior the limes $f_b \to 1$ and $f_b to 0$ should be similar. Instead, we see a monotonous dependence of $P/<P_{r}>$ which can be well fitted by a negative power-law
of the form $f_{b}^{\alpha}$, with $\alpha \sim 0.5$. The figures ~\ref{fig6}a) and b) indicate that when blacks are in a small minority, the frequency of the inter-ethnic relations correspond to an almost perfect desegregation, while in the other extreme, when whites are in a small minority, extremely strong segregation occurs. Our results suggest the following picture: Both whites and blacks show increasing homophily as their 
get into minorities. However, blacks as a small minority in a white majority get more integrated than the other way around. This result points toward the finding that the increase of racial heterogeneity
does not necessarily favor the inter-ethnic nominations among the increasing minority and the race of the majority, but may have the opposite effect~\cite{Moody}.

\section*{Conclusions}

In this article we have applied network concepts and tools to investigate the social structure of schools. We used the Add Health data base \cite{AddHealth} which contains - among others - detailed data about friendship nominations, race, age, gender, etc. We have first analyzed the weighted friendship network where the weight of a link between students $i \rightarrow j$ corresponds to the number of sheared activities of $i$ with $j$ as nominated by $i$. We have found striking asymmetries in the nominations and concluded that the community structures can be best uncovered if the underlying networks are chosen with the weakest criteria (one nomination in either direction already results in a link).

We have presented the statistical properties of these networks. The community structure was studied by means of $k$-clique percolation and the c-network of communities was constructed using overlap generated links. The optimal clique size was found to be $k=3$ in agreement with the special role of triads in social interactions.
While the friendship networks show the expected assortativity and their degree distribution have a sharp cutoff, the c-networks are disassortative and they have a scale-free degree distribution.

Finally, we presented a statistical analysis of ethnic preferences in friendship selection based on a comparison of the relative frequencies of $r-r'$ links as compared to a randomized reference system. We have analyzed the preference order of the four major ethnic groups. Furthermore we concluded that very small black minorities in a white majority have better balanced inter-ethnic relations than a small white minority in a black majority.
This could be related to the non-trivial effect of increasing ethnic heterogeneity on desegregation.

This research has been supported by grants OTKA TO49674 and K60456. MCG thanks DAAD for financial support.
HJH thanks the Max Planck Prize. 


\end{document}